\documentclass{article}

\usepackage[english]{babel}

\usepackage[a4paper,top=2cm,bottom=2cm,left=3cm,right=3cm,marginparwidth=1.75cm]{geometry}

\usepackage{amssymb}
\usepackage{siunitx}
\PassOptionsToPackage{hyphens}{url}\usepackage{hyperref}
\usepackage{cleveref}
\usepackage[utf8]{inputenc}
\usepackage[right]{lineno}
\usepackage{csquotes}
\usepackage{booktabs}
\usepackage{longtable}
\usepackage{adjustbox}
\usepackage{array}
\usepackage{url}
\usepackage{titlesec}
\usepackage{authblk}
\usepackage{xcolor} 

\titleformat{\subsection}
  {\mdseries\itshape\large} 
  {\thesubsection}{1em}{} 

\usepackage[english]{babel}
\usepackage[style=authoryear,backend=biber,natbib=true,maxcitenames=2,uniquelist=false]{biblatex}
\DeclareNameAlias{sortname}{family-given}
\DeclareNameAlias{default}{family-given}

\renewbibmacro{in:}{}
\DeclareFieldFormat[article]{title}{\mkbibquote{#1}\addcomma}
\DeclareFieldFormat[book]{title}{\mkbibemph{#1}\addcomma}
\DeclareFieldFormat[bookinbook]{title}{\mkbibemph{#1}\addcomma}
\DeclareFieldFormat[inbook]{title}{\mkbibquote{#1}\addcomma}
\DeclareFieldFormat[incollection]{title}{\mkbibquote{#1}\addcomma}
\DeclareFieldFormat[inproceedings]{title}{\mkbibquote{#1}\addcomma}
\DeclareFieldFormat[manual]{title}{\mkbibemph{#1}\addcomma}
\DeclareFieldFormat[misc]{title}{\mkbibemph{#1}\addcomma}
\DeclareFieldFormat[thesis]{title}{\mkbibemph{#1}\addcomma}
\DeclareFieldFormat[unpublished]{title}{\mkbibquote{#1}\addcomma}
\DeclareFieldFormat[patent]{title}{\mkbibemph{#1}\addcomma}
\DeclareFieldFormat[report]{title}{\mkbibemph{#1}\addcomma}
\DeclareFieldFormat[online]{title}{\mkbibquote{#1}\addcomma}
\DeclareFieldFormat[software]{title}{\mkbibemph{#1}\addcomma}
\DeclareFieldFormat[booklet]{title}{\mkbibemph{#1}\addcomma}
\DeclareFieldFormat[periodical]{title}{\mkbibemph{#1}\addcomma}
\DeclareFieldFormat[standard]{title}{\mkbibemph{#1}\addcomma}

\DeclareFieldFormat[article]{journaltitle}{\iffieldundef{shortjournal}{\mkbibemph{#1}\addcomma}{\mkbibemph{\printfield{shortjournal}}\addcomma}}
\DeclareFieldFormat{volume}{\bibstring{volume}~#1}
\DeclareFieldFormat{number}{\bibstring{number}~#1}

\DefineBibliographyStrings{english}{
  volume = {Vol.},
  number = {No.}
}

\renewbibmacro*{volume+number+eid}{%
  \printfield{volume}%
  \setunit*{\addspace}%
  \printfield{number}%
  \setunit{\addcomma\space}%
  \printfield{eid}}

\renewbibmacro*{journal+issuetitle}{%
  \usebibmacro{journal}%
  \setunit*{\addcomma\space}%
  \usebibmacro{volume+number+eid}%
  \setunit{\addcomma\space}%
  \usebibmacro{issue+date}}

\renewbibmacro*{publisher+location+date}{%
  \printlist{publisher}%
  \iflistundef{location}
    {\setunit*{\addcomma\space}}
    {\setunit*{\addcolon\space}}%
  \printlist{location}%
  \setunit*{\addcomma\space}%
  \usebibmacro{date}}

\DeclareCiteCommand{\cite}[\mkbibparens]
  {\usebibmacro{prenote}}
  {\usebibmacro{citeindex}%
   \usebibmacro{cite}}
  {\multicitedelim}
  {\usebibmacro{postnote}}

\renewbibmacro*{cite:labelyear+extrayear}{%
  \iffieldundef{labelyear}
    {}
    {\printtext[bibhyperref]{%
       \printfield{labelyear}%
       \printfield{extrayear}}}}

\renewbibmacro*{cite:labeldate+extradate}{%
  \iffieldundef{labelyear}
    {}
    {\printtext[bibhyperref]{%
       \printfield{labelyear}%
       \printfield{extradate}}}}

\AtEveryBibitem{
  \clearfield{month}
  \clearfield{day}
  \ifentrytype{book}{
    \clearlist{location}
  }{}
}

\DefineBibliographyStrings{english}{
  andothers = {\textit{et al.},}
}

\DeclareFieldFormat[article]{volume}{\bibstring{jourvol}\addnbspace #1}
\DeclareFieldFormat[article]{number}{\bibstring{number}\addnbspace #1}
\DeclareFieldFormat[article]{volume}{Vol. #1}
\DeclareFieldFormat[article]{number}{No. #1}

\DeclareFieldFormat{url}{\bibstring{available at}\addcolon\space\url{#1}}
\DeclareFieldFormat{urldate}{\mkbibparens{accessed \addspace#1}}

\DeclareFieldFormat{urldate}{%
  \mkbibparens{accessed\space%
    \thefield{urlday}\addspace%
    \mkbibmonth{\thefield{urlmonth}}\addspace%
    \thefield{urlyear}}}

\crefformat{figure}{#2Figure~#1#3}
\Crefformat{figure}{#2Figure~#1#3}
\crefformat{table}{#2Table~#1#3}
\Crefformat{table}{#2Table~#1#3}
\crefformat{section}{#2Section~#1#3}
\Crefformat{section}{#2Section~#1#3}

\author[1]{Nickolas Solomey}
\author[5]{Mark Christl}
\author[1]{Brian Doty}
\author[1]{Jonathan Folkerts}
\author[1]{Brooks Hartsock}
\author[4]{Evgen Kuznetsco}
\author[2]{Robert McTaggart}
\author[1]{Holger Meyer}
\author[1]{Tyler Nolan}
\author[3]{Greg Pawloski}
\author[1]{Daniel Reichart}
\author[5]{Miguel Rodriguez-Otero}
\author[1]{Dan Smith$^*$}
\author[1]{Lisa Solomey$^+$}

\affil[1]{Wichita State Univ., Department of Mathematics, Statistics and Physics, Kansas}
\affil[2]{South Dakota State University, Department of Physics, Brookings}
\affil[3]{Univ. of Minnesota, Department of Physics and Astronomy, Minneapolis}
\affil[4]{Univ. of Alabama at Huntsville, Department of Electrical Engineering}
\affil[5]{NASA Marshall Space Flight Center, Huntsville, Alabama}

\title{New methods of neutrino and anti-neutrino detection from 0.115 to 105 MeV \\
{\large Contribution to the 25th International Workshop on Neutrinos from Accelerators}}

\begin{document}
\maketitle

$\hspace{-5mm} ^*$From Hutchinson Community College as Watkins summer faculty visiting scholar. \\
$\; ^+$From Univ. of Chicago as a summer undergraduate research experience student.

\begin{abstract}
We have developed a neutrino detector with threshold energies from ~0.115 to 105 MeV in a clean detection mode almost completely devoid of accidental backgrounds. It was initially developed for the NASA $\nu$SOL project to put a solar neutrino detector very close to the Sun with 1,000 to 10,000 times higher solar neutrino flux than on Earth. A Similar detector could measure anti-neutrinos to study Beta decay neutrinos from reactors, geological sources, or for nuclear security applications. These techniques work at the 1 to 100 MeV region for neutrinos from the ORNL Spallation Neutron Source or low energy accelerator neutrinos and anti-neutrinos less than $\sim$100 MeV.  The identification process is clean, with a double pulse detection signature within a time window between the first interaction producing the conversion electron or positron and the secondary gamma emission of 100 ns to ~1 $\mu$s, which removes most accidental backgrounds. These new modes of detection could allow improvements to neutrino interaction measurements from an accelerator beam on a target. 
\end{abstract}

\section{Solar Neutrino Detection}
Neutrino detectors are normally large and deep underground. They are large because they need as much mass as possible to capture more neutrinos, and they are underground to reduce backgrounds. 
 
If a neutrino detector could operate in space \cite{1}, what advantages would it have to perform unique science? Going to space has the obvious application of approaching the sun, but several interesting possibilities arise in space besides the obvious option of approaching the sun. However, moving further from the sun would lower the solar neutrino floor, potentially allowing more sensitive dark matter searches.
 
We propose three distinct possibilities for science\cite{2}: 
\begin{enumerate}
\item Going closer to the Sun has the 1/r$^2$ neutrino flux increasing providing 1,000x higher neutrino flux at $7R_\odot$, where the current NASA Parker Solar probe operates. At $3R_\odot$, where some NASA scientists think it is possible to go, the neutrino flux increases by 10,000x . Here, new science could result from a larger statistical observation of solar neutrino emission that could enhance the study of the size of the fusion emission region, and particle physics looking for the transition from coherent to de-coherent neutrino oscillations as the neutrinos propagate away from the Sun. 
\item Because the neutrino has a non-zero mass, which we know from the existence of neutrino oscillations, the Sun bends space to create a gravitational focus. For neutrinos, this gravitational focus would be very close to us at only 20 to 40 AU. The galactic core is 25,000 light years away from us, is about two times larger than the moon when viewed from Earth, and is the 2nd largest neutrino source in the sky after the Sun. The galactic core not only has many neutrino-producing stars, but also $\sim$10,000 neutron stars and black holes in the central cubic parsec of the core. Matter falling into these objects is crushed, producing neutrons and isotopically emitted neutrinos of higher energy than solar fusion neutrinos. A detector at the solar neutrino focus would permit imaging of the galactic core using its 10$^{13}$ “light” collecting power and just finding the neutrino gravitational focus of the Sun would be a new way to measure the neutrino mass. 
\item Any neutrino-detecting space probe could take advantage of a flight towards the Sun to observe the shape of the solar neutrino changes due solely to the distance from a non-point source of solar neutrinos inside the Sun, while when traveling away from the Sun, deviations from the expected 1/r$^2$ curve are an indication of the direct observation of Dark Matter or galactic neutrinos.
\end{enumerate}
We have devised a technique to observe solar neutrinos in space using a double delayed coincidence pulse within a $\sim$250 ns time window \cite{2,3}, see figure 1-left. The double pulse adds a clean identification method while losing only a fraction of the solar neutrino interactions\cite{4}.

\begin{figure}[h]
 \centering
 \makebox[\textwidth][c]{\includegraphics[width=1\textwidth]{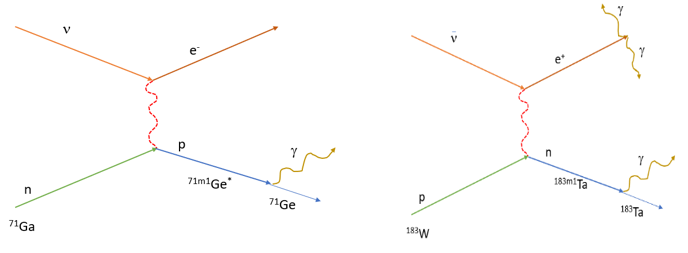}}%
 \caption{On the left is a diagram for a Solar neutrinos from Hydrogen fusion in the sun’s core captured on $^{71}$Ga Isotope producing a delayed secondary pulse. On the right is a similar process for anti-neutrinos using $^{183}$W but other options like $^{107}$Ag are possible, see Table 1.}
 \label{fig-1}
\end{figure}

\begin{table}[h]
 \centering
 \makebox[\textwidth][c]{\includegraphics[width=1\textwidth]{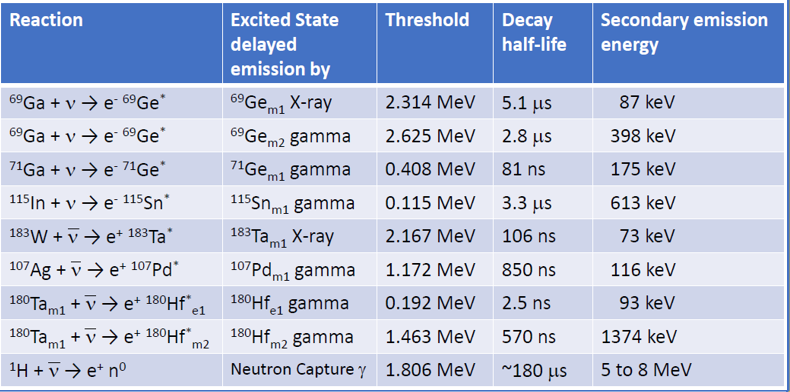}}%
 \caption{Properties of all the double pulse with fast decay isotopes for neutrino and anti-neutrino interactions such as threshold, secondary decay half-life, and secondary decay gamma ray emission energy. Although the $^{71}$Ga mode has been seen, its cross section has not been measured and the neutrino interaction on $^{115}$In and the anti-neutrino interaction on $^{183}$W, $^{107}$Ag and $^{180}$Ta have not been seen but are certain to exist.}
 \label{tab-II}
\end{table}

\section{Reactor Anti-Neutrino Detection}
Our group has developed a nuclear reactor anti-neutrino detector, using a similar anti-neutrino reaction process to our Solar Neutrino detection method. We initially focused on $^{183}$W \cite{5} with a double delayed coincidence pulse as seen in figure 1-right. We have since discovered other fast processes on $^{107}$Ag and $^{180}$Ta which have thresholds from 0.19 MeV to 2 MeV. An interesting application of this anti-neutrino detector is that, with a large enough detector, it could possibly detect submersed and highly movable fission reactors. Furthermore, we propose to build a test detector prototype using $^{183}$W or $^{107}$Ag anti-neutrino interactions which would be a proof-of-concept demonstrator. The detector would be operated either at the HFIR 85 MW ORNL reactor or the 10 MW Missouri University Research Reactor. Both of these detectors have well-understood anti-neutrino fluxes, and operations at either would allow for a measurement of the absolute cross section for the target element.

The goal of a small prototype anti-neutrino and solar neutrino detector is threefold:
\begin{enumerate}
\item To observe the delayed coincidence pulses of solar neutrino type interactions on Ga transmuting to a nuclear excited state of Ge and measure its cross-section. Both the $<$ms process and the $\sim$100 ns process could potentially be used to detect and measure solar neutrinos when a detector is close to the Sun with minimal shielding with a Gallium-Aluminum-Gadolinium Garnet (GAGG) scintillator detector surrounded by an active veto array. 
\item A corollary of this project has been a interest in developing an improved anti-neutrino detector. Fission reactors such as those on submarines are a continuous source of anti-neutrinos, even if the reactor is off.  Using simulations, we have studied the nuclear isotope materials for such a detector using the standard double pulse Inverse Beta Decay (IBD) technique first used by Cowan and Reines in the 1950s. This produces a conversion positron with an emitted neutron where the neutron can then be absorbed by an isotope with a large neutron capture cross section and its subsequent gamma emissions. We have also developed a double pulse technique like the Ga to Ge* process being used in our NASA project, but for anti-neutrinos by using the isotope $^{183}$W transition to $^{183}$Ta* where the nuclear excited state of 107 ns half-life emits a delayed 73 keV gamma to reduce background and with a minimal amount of shielding and an active charged particle veto. 
\item The red giant star Betelgeuse may go supernova within decades. Recent peer reviewed articles have shown that Helium burning in this star is complete, and the star is burning carbon and sulfur. A supernova can produce a tremendous number of neutrinos over a six-second period, as was seen with the 1987 supernova. That supernova was 170,000 light years away, and an equivalent supernova of Betelgeuse or another nearby star would result in two million events in the same size detector over six seconds. Because a nearby supernova would provide a higher neutrino flux at earth, even a small detector can find a nearby supernova. Measuring the neutrino and anti-neutrino supernova flux with a small detector would be a notable scientific achievement. Because our detection technique can distinguish lower energies, it would be able to see the neutrino time evolution of the supernova pulse. This prototype test detector for ORNL could eventually become, in a Salt Mine of Kansas 650 ft below ground, a nice monitoring station once neutrino and anti-neutrino detection is calibrated at the ORNL HFIR and SNS test areas.
\end{enumerate}

\section{Accelerator Beamline Neutrino production studies}
If the momentum transfer from the neutrino interaction to an excited state nucleus is below the threshold for disintegrating that nucleus, then this technique could be used to study neutrino and anti-neutrino production from a target up to the muon production threshold of 105 MeV/c, see figure 2.

\begin{figure}[h]
 \centering
 \makebox[\textwidth][c]{\includegraphics[width=1\textwidth]{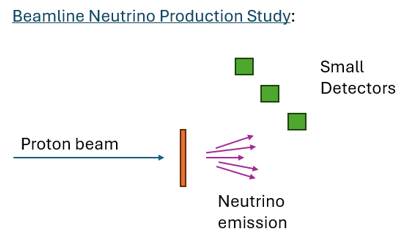}}%
 \caption{An array of small detectors are various angles to study neutrino and anti-neutrino production off a target in the beam-line.}
 \label{fig-2}
\end{figure}

\section{Conclusion}
In collaboration with ORNL it is hoped that the DOE would fund our study for a small prototype demonstrator detector for neutrinos at the ORNL Spallation Neutron Source and for the anti-neutrino methods at either HFIR or MURR reactor. Once demonstrated the small detector could be put downstream of a target being hit by protons at various angles to look at neutrino and anti-neutrino production in the MeV regime.

\section*{Acknowledgments} 
This work was supported by the NASA STMD NIAC Program [grant numbers 80NSSC2K1900, 80NSSC18K0868, 80NSSC19M0971]; MSFC CAN18 and CAN24 [grant numbers 80MSFC18M0047 and 80NSSC24M0059]; NASA Heliophysics SMD HITS grant number 80NSSC24K0070; NASA Jump Start program in Kansas grant 80NSSC20M0109, internal Wichita State University MURPA; and NASA in Kansas EPSCoR Partnership Development Grant.

\end{document}